\PassOptionsToPackage{unicode}{hyperref}
\PassOptionsToPackage{hyphens}{url}
\documentclass[
  11pt,
]{article}
\usepackage{xcolor}
\usepackage[margin=1in]{geometry}
\usepackage{amsmath,amssymb}
\usepackage{iftex}
\ifPDFTeX
  \usepackage[T1]{fontenc}
  \usepackage[utf8]{inputenc}
  \usepackage{textcomp} % provide euro and other symbols
\else % if luatex or xetex
  \usepackage{unicode-math} % this also loads fontspec
  \defaultfontfeatures{Scale=MatchLowercase}
  \defaultfontfeatures[\rmfamily]{Ligatures=TeX,Scale=1}
\fi
\usepackage{lmodern}
\ifPDFTeX\else
\fi
\IfFileExists{upquote.sty}{\usepackage{upquote}}{}
\IfFileExists{microtype.sty}{% use microtype if available
  \usepackage[]{microtype}
  \UseMicrotypeSet[protrusion]{basicmath} % disable protrusion for tt fonts
}{}
\makeatletter
\@ifundefined{KOMAClassName}{% if non-KOMA class
  \IfFileExists{parskip.sty}{%
    \usepackage{parskip}
  }{% else
    \setlength{\parindent}{0pt}
    \setlength{\parskip}{6pt plus 2pt minus 1pt}}
}{% if KOMA class
  \KOMAoptions{parskip=half}}
\makeatother
\usepackage{longtable,booktabs,array}
\usepackage{calc} % for calculating minipage widths
\usepackage{etoolbox}
\makeatletter
\patchcmd\longtable{\par}{\if@noskipsec\mbox{}\fi\par}{}{}
\makeatother
\IfFileExists{footnotehyper.sty}{\usepackage{footnotehyper}}{\usepackage{footnote}}
\makesavenoteenv{longtable}
\usepackage{graphicx}
\makeatletter
\newsavebox\pandoc@box
\newcommand*\pandocbounded[1]{% scales image to fit in text height/width
  \sbox\pandoc@box{#1}%
  \Gscale@div\@tempa{\textheight}{\dimexpr\ht\pandoc@box+\dp\pandoc@box\relax}%
  \Gscale@div\@tempb{\linewidth}{\wd\pandoc@box}%
  \ifdim\@tempb\p@<\@tempa\p@\let\@tempa\@tempb\fi% select the smaller of both
  \ifdim\@tempa\p@<\p@\scalebox{\@tempa}{\usebox\pandoc@box}%
  \else\usebox{\pandoc@box}%
  \fi%
}
\def\fps@figure{htbp}
\makeatother
\providecommand{\tightlist}{%
  \setlength{\itemsep}{0pt}\setlength{\parskip}{0pt}}
\usepackage{graphicx}
\usepackage{float}
\usepackage{xurl}
\usepackage{newunicodechar}
\usepackage{amssymb}
\usepackage{textcomp}

\usepackage{fvextra}
\fvset{breaklines=true, breakanywhere=true, breaksymbol={}}
\RecustomVerbatimEnvironment{verbatim}{Verbatim}{breaklines=true,breakanywhere=true,breaksymbol={}}

\usepackage{microtype}

\usepackage{etoolbox}
\AtBeginEnvironment{longtable}{\small}

\usepackage{caption}

\usepackage{iftex}

\ifPDFTeX

\newunicodechar{ℝ}{\ensuremath{\mathbb{R}}}

\newunicodechar{⁶}{\textsuperscript{6}}
\newunicodechar{ⁿ}{\textsuperscript{n}}
\newunicodechar{ᵀ}{\textsuperscript{T}}

\newunicodechar{₀}{\textsubscript{0}}
\newunicodechar{₁}{\textsubscript{1}}
\newunicodechar{₂}{\textsubscript{2}}
\newunicodechar{₃}{\textsubscript{3}}
\newunicodechar{₄}{\textsubscript{4}}
\newunicodechar{₅}{\textsubscript{5}}
\newunicodechar{₆}{\textsubscript{6}}
\newunicodechar{₇}{\textsubscript{7}}
\newunicodechar{₈}{\textsubscript{8}}
\newunicodechar{₉}{\textsubscript{9}}

\newunicodechar{ₙ}{\textsubscript{n}}
\newunicodechar{ₖ}{\textsubscript{k}}
\newunicodechar{ₜ}{\textsubscript{t}}
\newunicodechar{ᵢ}{\textsubscript{i}}
\newunicodechar{ⱼ}{\textsubscript{j}}

\newunicodechar{₊}{\textsubscript{+}}
\newunicodechar{₋}{\textsubscript{-}}

\newunicodechar{−}{\ensuremath{-}}
\newunicodechar{∈}{\ensuremath{\in}}
\newunicodechar{∉}{\ensuremath{\notin}}
\newunicodechar{≤}{\ensuremath{\leq}}
\newunicodechar{≥}{\ensuremath{\geq}}
\newunicodechar{≠}{\ensuremath{\neq}}
\newunicodechar{≈}{\ensuremath{\approx}}
\newunicodechar{→}{\ensuremath{\rightarrow}}
\newunicodechar{←}{\ensuremath{\leftarrow}}
\newunicodechar{×}{\ensuremath{\times}}
\newunicodechar{∑}{\ensuremath{\sum}}
\newunicodechar{∞}{\ensuremath{\infty}}

\newunicodechar{ⓘ}{\textcircled{\scriptsize i}}

\newunicodechar{θ}{\ensuremath{\theta}}
\newunicodechar{μ}{\ensuremath{\mu}}
\newunicodechar{σ}{\ensuremath{\sigma}}
\newunicodechar{τ}{\ensuremath{\tau}}
\newunicodechar{φ}{\ensuremath{\varphi}}
\newunicodechar{α}{\ensuremath{\alpha}}
\newunicodechar{β}{\ensuremath{\beta}}
\newunicodechar{ε}{\ensuremath{\varepsilon}}
\newunicodechar{λ}{\ensuremath{\lambda}}
\newunicodechar{γ}{\ensuremath{\gamma}}
\newunicodechar{δ}{\ensuremath{\delta}}
\newunicodechar{ρ}{\ensuremath{\rho}}
\newunicodechar{π}{\ensuremath{\pi}}
\newunicodechar{ω}{\ensuremath{\omega}}

\fi
\usepackage{bookmark}
\IfFileExists{xurl.sty}{\usepackage{xurl}}{} % add URL line breaks if available
\hypersetup{
  hidelinks,
  pdfcreator={LaTeX via pandoc}}

\author{}
\date{}

\begin{document}

\section{Pramāṇa: A Protocol-Layer Treatment of Claim Verification in
Autonomous Agent
Networks}\label{pramux101ux1e47a-a-protocol-layer-treatment-of-claim-verification-in-autonomous-agent-networks}

\textbf{Ravi Kiran Kadaboina}

\emph{Independent Researcher}

\emph{M.S., Computer Engineering, University of New Mexico, 2011}

\subsection{Abstract}\label{abstract}

Autonomous agents deployed in regulated domains must produce a
verification artifact per consequential output. The artifact is a record
an auditor can re-execute offline, capturing what was claimed, against
what source, by whom, when, and how. Production verification today
splits into two unstandardized halves. Probabilistic verdict patterns
(self-consistency voting, confidence-scored outputs, reviewer LLM (large
language model) ensembles) produce judgments about model outputs, not
artifacts. Artifact-producing patterns (retrieval-augmented generation
with citations, tool-augmented traces, generator-verifier loops, and
recent multi-agent research systems such as Google's AI co-scientist
{[}42{]}) produce vendor-specific records that an external auditor
cannot reconstruct without bespoke integration. Pramana defines the
missing wire format: a typed claim attestation with a deterministic
\texttt{verify()} contract, layered as an A2A and MCP extension.

Pramana wraps every consequential agent output in a typed
\texttt{ClaimAttestation} with one of four variants (measurement,
inference, analogy, citation), each paired with a \texttt{verify()}
operation against the recorded source. For \texttt{MeasurementClaim} and
\texttt{CitationClaim}, \texttt{verify()} is a deterministic function of
\texttt{(claim,\ source)} and no probabilistic judge participates in the
verification step. For \texttt{InferenceClaim} and \texttt{AnalogyClaim}
the determinism is conditional on the oracle: deterministic when the
oracle is a theorem prover, structural matcher, or fixed-metric
similarity function; audit-replayable when the oracle is an LLM. The
four-way typology is derived from classical Indian epistemology
(pramana, ``the means of valid knowledge'') and partitions agent claims
by their ground.

The Pramana lifecycle is specified in TLA+ (Temporal Logic of Actions)
and exhaustively verified under TLC (TLA+'s model checker) across three
symmetry-reduced models, totaling 38,563 distinct reachable states with
zero invariant violations. The Python reference implementation passes 84
unit and property-based tests. A wire-extension manifest publishes the
format for A2A (Agent2Agent) {[}1{]} and MCP (Model Context Protocol)
{[}2{]} agents, with three deployment-grade invariants (reachability,
service-level-agreement (SLA) bound, and offline re-verifiability)
layered on top.

We report an exploratory pilot (n=100 problems, approximately 1.3M
tokens across 2,275 reviewer calls) that probes the LLM-as-judge
alternative in code generation. The strongest observation is a roughly
40-percentage-point raw FPR (false-positive rate) delta between corpora
under the same ensemble, consistent with reference-solution quality
contributing significantly; differential prompt-overfit across corpora
during the selection phase is an alternative driver that the
single-rater design cannot rule out. The pilot does not validate Pramana
on its own; the structural argument in Sections 1 through 4 and the
formal verification in Section 6 do that.

\begin{center}\rule{0.5\linewidth}{0.5pt}\end{center}

\subsection{1. Introduction}\label{introduction}

A high-stakes agent deployment needs more than an output that is
probably correct. It needs a record of what was claimed, against what
source, by whom, when, in a form an external auditor can re-execute
offline. Production verification today splits into two halves.
Probabilistic verdict patterns (self-consistency voting, confidence
scores, reviewer LLM ensembles) produce judgments about model outputs,
not artifacts; aggregating judgments does not change their category.
Artifact-producing patterns (retrieval-augmented generation with
citations, tool-augmented traces, generator-verifier loops as in
FunSearch {[}30{]} and AlphaEvolve {[}31{]}, and recent multi-agent
research systems such as Google's AI co-scientist {[}42{]}) produce
vendor-specific records but no shared wire format. An auditor inspecting
a multi-vendor agent network cannot reconstruct verification across
patterns without bespoke per-vendor integration.

The verification artifact a regulator wants has a structurally different
shape from a verdict. An artifact records the claim, the source it was
checked against, the procedure used to check it, the verifier identity,
the timestamp, and the outcome. Pramana defines the typed wire format
that standardizes this artifact. It is published as an A2A and MCP
extension manifest at
\texttt{https://ravikiran438.github.io/pramana-attestation/v1} via the
\texttt{capabilities.extensions} mechanism those host protocols define,
alongside sibling extensions (Anumati, Yathartha, Phala, Sauvidya,
Pratyahara) addressing other wire-level concerns.

The rest of this paper is organized as follows. Section 1.1 describes
what the protocol layer is currently missing. Section 1.2 surveys the
regulatory pressure motivating this work. Section 1.3 lists the
contributions. Section 2 defines the Pramana primitive. Section 3
describes composition with the existing stack. Section 4 maps the
regulatory landscape onto specific primitives. Section 5 reports the
supplementary empirical pilot. Section 6 reports the formal correctness
properties. Section 7 surveys related work. Section 8 lists limitations
and future work.

\subsubsection{1.1 What the protocol layer is
missing}\label{what-the-protocol-layer-is-missing}

A2A {[}1{]} and MCP {[}2{]} standardize agent communication at the
syntactic level (messages, tasks, agent cards, tool calls, tool
results), but neither defines a typed vocabulary for the \emph{epistemic
ground} of an agent output. A tool result in MCP is a content blob; an
A2A agent message is a text payload. Neither carries a typed attestation
of source URI, measurement record, or inference chain. Pramana adds the
missing vocabulary, with each variant naming the epistemic ground that
warrants the claim plus the metadata required to verify it
independently.

\texttt{ClaimAttestation} is parameterized by claim type. For
\texttt{MeasurementClaim} and \texttt{CitationClaim} the verification
step is structurally deterministic: \texttt{verify()} is a function of
\texttt{(claim,\ source)} and no probabilistic judge participates. For
\texttt{InferenceClaim} and \texttt{AnalogyClaim} the verification
step's determinism is conditional on the oracle plugged into the
\texttt{verify()} slot, deterministic when the oracle is a theorem
prover, structural matcher, or fixed-metric similarity function, and
audit-replayable rather than bit-deterministic when the oracle is an
LLM. Section 2.5 specifies the partition.

\subsubsection{1.2 Regulatory pressure}\label{regulatory-pressure}

Regulators across jurisdictions converge on artifact-requiring
compliance bars. CFPB Circular 2023-03, Federal Reserve SR 11-7, NYDFS
Insurance Circular Letter No.~7, HHS OCR's Section 1557 Final Rule, the
EU AI Act, NYC Local Law 144, and the Colorado and California state AI
laws all require per-decision documentary records that an external
auditor can re-execute. The regulatory frameworks specify the audit
substrate (a re-verifiable record per consequential output) rather than
a particular verification pattern. Section 4 maps each requirement to
the Pramana primitive that produces the substrate.

\subsubsection{1.3 Contributions}\label{contributions}

\begin{enumerate}
\def\labelenumi{\arabic{enumi}.}
\tightlist
\item
  A typed wire-format primitive: a \texttt{ClaimAttestation} with four
  variants (measurement, inference, analogy, citation) derived from the
  classical pramana taxonomy of valid knowledge, each paired with a
  \texttt{verify()} operation against the recorded source. The four
  variants partition agent claims by epistemic ground; verification
  determinism by claim type is specified in Section 2.5.
\item
  A TLA+ specification of the lifecycle, exhaustively verified under TLC
  across three symmetry-reduced models (38,563 distinct reachable
  states, zero invariant violations), plus a Python reference
  implementation that passes 84 unit and property-based tests.
\item
  An A2A and MCP wire-extension manifest (\texttt{claim-attestation/v1})
  with three deployment-grade invariants: reachability, SLA-bound, and
  offline re-verifiability.
\item
  An exploratory pilot (n=100, approximately 1.3M tokens) that probes
  the LLM-as-judge alternative in code generation and characterizes the
  measurement regress in reviewer-ensemble evaluation.
\end{enumerate}

\begin{center}\rule{0.5\linewidth}{0.5pt}\end{center}

\subsection{2. The Pramana Primitive}\label{the-pramana-primitive}

A \texttt{ClaimAttestation} is a structured assertion an agent makes
about something it has observed, inferred, compared, or cited. Each
attestation declares the epistemic ground that warrants the claim plus
enough metadata for any party holding the source to verify the claim
independently.

\subsubsection{2.1 The four claim types}\label{the-four-claim-types}

The taxonomy follows the four ways a claim can be grounded in observable
reality.

{\def\LTcaptype{none} % do not increment counter
\begin{longtable}[]{@{}
  >{\raggedright\arraybackslash}p{(\linewidth - 4\tabcolsep) * \real{0.3333}}
  >{\raggedright\arraybackslash}p{(\linewidth - 4\tabcolsep) * \real{0.3333}}
  >{\raggedright\arraybackslash}p{(\linewidth - 4\tabcolsep) * \real{0.3333}}@{}}
\toprule\noalign{}
\begin{minipage}[b]{\linewidth}\raggedright
Type
\end{minipage} & \begin{minipage}[b]{\linewidth}\raggedright
Ground
\end{minipage} & \begin{minipage}[b]{\linewidth}\raggedright
Verification operation
\end{minipage} \\
\midrule\noalign{}
\endhead
\bottomrule\noalign{}
\endlastfoot
\texttt{MeasurementClaim} & Direct observation or structured record &
Source-record fetch and field-match \\
\texttt{InferenceClaim} & Logical inference from prior claims &
Inference-chain replay \\
\texttt{AnalogyClaim} & Similarity to a known reference case &
Similarity recompute against reference \\
\texttt{CitationClaim} & Attribution to an authoritative source & Source
fetch plus faithful-citation check \\
\end{longtable}
}

Each subtype is a Pydantic v2 model in a discriminated union over a
\texttt{claim\_type} field. Pydantic v2 is the data-modeling library
used by both the A2A {[}1{]} and MCP {[}2{]} reference SDKs, so a
Pramana extension deserializes through the same code path the host agent
already runs for native A2A or MCP messages. All four share a common
envelope (\texttt{claim\_id}, \texttt{claim\_type}, \texttt{claim},
\texttt{attester\_id}, \texttt{attested\_at}) and contribute
type-specific fields: a \texttt{MeasurementClaim} adds the measured
value, unit, method, source URI, and uncertainty; a
\texttt{CitationClaim} adds the source URI, the verbatim excerpt, and a
retrieval timestamp. Full per-type schemas are in the companion
repository.

\subsubsection{2.2 Verification semantics}\label{verification-semantics}

Each variant's \texttt{verify()} is a function from
\texttt{(claim,\ source)} to one of three outcomes: \texttt{VERIFIED},
\texttt{REJECTED}, or \texttt{UNVERIFIABLE}. The function is
deterministic when its injected dependency is deterministic, such that
re-running with the same inputs produces a bit-identical
\texttt{VerificationOutcome} (up to timestamp and verifier id). Source
fetchers and hash matchers are naturally deterministic; LLM-backed
inference oracles and similarity computers are not, and Section 2.5
spells out the audit-replayability fallback. The function is also
injectable, with source-fetch, measurement-fetch,
similarity-computation, and inference- oracle dependencies as
constructor arguments. This permits production-grade caching, SSRF
(server-side request forgery) protection, allow-listing, and offline
auditing without changing the verification logic.

The \texttt{AnalogyClaim} case has one design subtlety worth surfacing.
When \texttt{similarity\_score} is recorded, \texttt{verify()} checks
the recomputed score against it within tolerance. When
\texttt{similarity\_score} is absent, \texttt{verify()} returns
\texttt{VERIFIED} as long as the similarity computer produced any score,
such that the deterministic check reduces to ``the similarity
computation succeeded.'' Emitters that want stricter verification should
record a \texttt{similarity\_score} explicitly. Regulated deployments
SHOULD require \texttt{similarity\_score} for any \texttt{AnalogyClaim}
and SHOULD reject VERIFIED outcomes lacking it. A VERIFIED outcome
without \texttt{similarity\_score} is documentary (the similarity
computation succeeded) rather than substantive (the similarity exceeded
a threshold), and an auditor reviewing such an outcome cannot
distinguish a successful match from a near-miss from a wire-format
inspection alone. The v1 wire format admits both modes intentionally to
support deployments without natural thresholds; a future revision could
introduce a distinct \texttt{VERIFIED\_UNTHRESHOLDED} status to make the
distinction visible at the audit layer.

\subsubsection{2.3 Lifecycle invariants}\label{lifecycle-invariants}

Five named safety invariants hold across all reachable states.

\begin{enumerate}
\def\labelenumi{\arabic{enumi}.}
\tightlist
\item
  \textbf{P-1 (Single Emission)}: each claim has at most one emission
  audit entry.
\item
  \textbf{P-2 (Verification Determinism)}: each claim's verification
  state is single-valued and reaches a terminal state at most once.
\item
  \textbf{P-3 (Audit Completeness)}: every terminal verification,
  suppression, and disclosure is recorded in the audit trail.
\item
  \textbf{P-4 (Disclosure Coupling)}: a claim shown to a principal has
  an \texttt{emitted}, a \texttt{verified}, and a \texttt{disclosed}
  entry in the audit trail.
\item
  \textbf{SuppressionDisclosureDisjoint}: no claim is both suppressed
  and shown to a principal.
\end{enumerate}

Full TLC verification is in Section 6.

\subsubsection{2.4 What Pramana verifies and what it does
not}\label{what-pramana-verifies-and-what-it-does-not}

Pramana is not a hallucination detector, a confidence score, or a
replacement for unit tests or human review. It names the verification
record that a regulator can demand and the function that produces it.
Whether the underlying agent is well-behaved is out of scope. Pramana
ensures that if the agent makes a claim, the claim is accompanied by a
record any auditor can re-execute.

The artifact-versus-judgment distinction is load-bearing for the
regulated-domain use case. Pramana does not propose to deterministically
verify the underlying judgment in subjective-ground-truth domains.
Whether a diagnosis is correct, whether a loan applicant is
creditworthy, whether an underwriting decision is fair: these are not
deterministic questions and Pramana does not pretend otherwise. What
Pramana deterministically verifies is the \emph{artifact} of the
decision: whether the \texttt{MeasurementClaim}'s reported lab value
matches the structured record, whether the \texttt{InferenceClaim}'s
named decision rule was applied as documented, whether the
\texttt{CitationClaim}'s reference to a guideline resolves to the cited
text, whether the \texttt{AnalogyClaim}'s stated similarity to a prior
case recomputes within tolerance. The regulator does not need to verify
the diagnosis; they need to verify the documentary record of how the
diagnosis was constructed.

\subsubsection{2.5 Threat model}\label{threat-model}

Pramana assumes the source registry is honest within the deployment's
trust boundary, the cryptographic primitives backing
\texttt{source\_digest} and \texttt{artifact\_signature} are not broken,
and verifiers run uncompromised code. Under these assumptions the
protocol defends against silent claim modification (CA-3 makes tampering
observable), against unilateral retraction (P-1 admits at most one
emission per claim), and against unbounded verification latency (CA-2
forces a terminal state within the declared SLA window). Pramana does
not defend against an adversary who fabricates source content at a URI
it controls; that attack is observable under Pramana but not prevented.
In deployments where source authority matters, Pramana composes with
separate source-attestation primitives (cryptographic signatures,
content-addressed storage, federated registries) via the
\texttt{source\_uri} integration point. Concretely, the
\texttt{source\_digest} field (Section 3.2) carries a cryptographic hash
of the source bytes at retrieval time; \texttt{CitationClaim} verifiers
MUST reject claims whose freshly-fetched content does not hash to the
recorded digest. This binding is mandatory for non-\texttt{unverifiable}
outcomes (CA-3) and is the protocol-level integration surface for
external source- attestation infrastructure. The protocol does not
specify particular attestation mechanisms, but constrains their
composition through the digest invariant.

The determinism guarantee holds unconditionally for
\texttt{MeasurementClaim} and \texttt{CitationClaim}, where the injected
dependency is a source fetcher or hash matcher. For
\texttt{InferenceClaim} and \texttt{AnalogyClaim} the guarantee holds
when the oracle is itself deterministic (theorem prover, structural
matcher, fixed-metric similarity function). When the oracle is a
commercial LLM API call, the guarantee fails. Modern hosted LLMs do not
produce bit-identical outputs even at temperature 0 and a fixed seed,
due to floating-point non-determinism in batched GPU inference, sparse
mixture-of-experts routing variability, and silent backend updates that
are not version-pinned by the provider. For these deployments Pramana's
contribution narrows to audit-replayability: the audit trail records the
oracle's inputs, its output, the model identifier, and the timestamp, so
an auditor can re-execute and observe both the original verdict and any
re-execution verdict that differs. Strict offline re-verifiability
(CA-3) requires constraining oracles to deterministic procedures.

Audit-replayability is categorically different from raw logging. The
trail captures \emph{structured} inputs (premises as a typed list,
conclusion as a separate field, \texttt{inference\_method} as a named
procedure, \texttt{source\_uri} as a resolvable reference), not the LLM
prompt blob. An auditor querying the trail can recover which decision
rule was claimed to apply, what premises were cited, and which sources
were consulted, independent of whether the oracle's verdict on
re-execution is bit-identical. Re-execution with the same oracle and
inputs surfaces the direction and magnitude of any divergence, which is
itself queryable evidence. Deployments that need full verdict
reproducibility plug in a deterministic oracle in the same dependency
slot, without changing the wire format or the audit-trail shape.

Composition with the existing agent-protocol stack, the wire-level
extensions to A2A and MCP, and the regulatory landscape this primitive
is designed to satisfy are addressed in the next three sections.

\begin{center}\rule{0.5\linewidth}{0.5pt}\end{center}

\subsection{3. Composition with the existing
stack}\label{composition-with-the-existing-stack}

\subsubsection{3.1 Composition with sibling
protocols}\label{composition-with-sibling-protocols}

Pramana is one of six wire-level extensions in the agent-protocol-stack
programme, each published as an A2A and MCP extension via the
\texttt{capabilities.extensions} mechanism the host protocols define.
Anumati / ACAP {[}10{]} specifies consent and adherence, where ACAP
records the consent for an action and Pramana records the verification
of each claim that fed the action. Yathartha {[}11{]} specifies the
capability surface, such that a capability self-attestation can be a
\texttt{CitationClaim} (weak grounding, with the agent itself as source)
and a third-party probed capability is a \texttt{MeasurementClaim}
(strong grounding). Phala {[}12{]} (welfare feedback), Pratyahara /
NERVE {[}13{]} (sensory integrity), and Sauvidya / PACE {[}14{]}
(accessibility envelope) compose adjacently. Each protocol follows the
same convention and Pramana takes the role of artifact verification in
the stack.

\subsubsection{3.2 Wire-level extension to
A2A}\label{wire-level-extension-to-a2a}

An agent declares Pramana support by including an entry in its A2A
AgentCard \texttt{capabilities.extensions} array whose \texttt{uri}
matches the Pramana extension URI declared in §1. The entry deserializes
to a \texttt{PramanaServiceRef} declaring the verify endpoint, audit
endpoint, supported \texttt{ClaimType} values, source-fetcher schemes,
signature algorithm, and SLA window. The receiving agent uses this
declaration to decide what claim types it can verify against and to
route incoming verification requests. The \texttt{claim-attestation/v1}
wire extension layers on top, adding a per-claim envelope with
\texttt{verify\_endpoint\_hint}, \texttt{verification\_artifact\_id},
\texttt{artifact\_signature}, \texttt{signature\_alg},
\texttt{source\_digest}, and \texttt{sla\_window\_ms}. The extension
enforces three deployment-grade invariants beyond the core lifecycle
invariants of Section 2.3. CA-1 (reachability) requires every emitted
attestation to carry a \texttt{verify\_endpoint\_hint} so any receiver
can re-verify. CA-2 (SLA bound) requires verification to reach a
terminal state within the declared window or be auto-marked
\texttt{unverifiable}. CA-3 (offline re-verifiability) requires every
non-\texttt{unverifiable} outcome to carry a \texttt{source\_digest} and
an \texttt{artifact\_signature} so an offline auditor can re-verify.

\subsubsection{3.3 Wire-level extension to
MCP}\label{wire-level-extension-to-mcp}

For MCP, \texttt{ClaimAttestation} records attach as tool-result
metadata under a \texttt{pramana} key in the \texttt{structuredContent}
field. The content payload is unchanged, and the attestation rides
alongside as structured metadata that an MCP-aware client can verify
before surfacing the content to the principal. Tools that do not
understand Pramana ignore the annotation, and tools that do can chain
attestations across tool calls without re-parsing the content payload.
The MCP wire encoding follows the same discriminated-union pattern as
the A2A wire format. Full schemas are in the companion repository.

The regulatory frameworks that require this artifact are surveyed next,
with the primitive-to-requirement mapping in Section 4.

\begin{center}\rule{0.5\linewidth}{0.5pt}\end{center}

\subsection{4. Regulatory landscape}\label{regulatory-landscape}

Regulators across jurisdictions converge on a common structure: per
consequential decision, an auditor must be able to recover what was
claimed, against what source, by whom, when, and how. None of the
frameworks below specify a particular mechanism. They specify the
verification record. Pramana provides one mechanism that produces it.
Table A summarizes the alignment; the framework details follow.

{\def\LTcaptype{none} % do not increment counter
\begin{longtable}[]{@{}
  >{\raggedright\arraybackslash}p{(\linewidth - 4\tabcolsep) * \real{0.3333}}
  >{\raggedright\arraybackslash}p{(\linewidth - 4\tabcolsep) * \real{0.3333}}
  >{\raggedright\arraybackslash}p{(\linewidth - 4\tabcolsep) * \real{0.3333}}@{}}
\toprule\noalign{}
\begin{minipage}[b]{\linewidth}\raggedright
Framework
\end{minipage} & \begin{minipage}[b]{\linewidth}\raggedright
Required record
\end{minipage} & \begin{minipage}[b]{\linewidth}\raggedright
Pramana primitive
\end{minipage} \\
\midrule\noalign{}
\endhead
\bottomrule\noalign{}
\endlastfoot
CFPB Circular 2023-03 / Reg B {[}3{]} & Specific source-linked reason
for adverse credit action & \texttt{VerificationOutcome.evidence} per
claim \\
SR 11-7 / OCC Bulletin 2011-12 {[}4{]} & Model documentation sufficient
for examiner re-execution & \texttt{ClaimAttestation} plus
\texttt{VerificationOutcome} per model output \\
NYDFS Circular Letter 7 (2024) {[}5{]} & Demonstrable rational
relationship for AI underwriting variables &
\texttt{InferenceClaim.verify()} against named premises \\
Colorado DOI Reg 10-1-1 {[}6{]} & Governance framework and per-decision
audit log for insurer AI & \texttt{ClaimAttestation} audit chain \\
HIPAA §164.312(b) & Examination of activity in ePHI systems &
\texttt{VerificationOutcome} per claim, append-only \\
Section 1557 / 45 C.F.R. §92.210 {[}7{]} & Evidence of ``reasonable
efforts'' to mitigate AI-tool discrimination & Re-executable
\texttt{VerificationOutcome} chain \\
EU AI Act Articles 14, 50 {[}8{]} & Effective human oversight and
decision disclosure & Inspectable \texttt{ClaimAttestation} per decision
component \\
GDPR Recital 71 {[}9{]} & Right to explanation & Per-claim source URI
plus excerpt plus re-verification procedure \\
\end{longtable}
}

\textbf{Table A.} Regulatory record requirements and the Pramana
primitive that produces each.

The mapping in Table A assumes claim types are paired with deterministic
oracles; for \texttt{InferenceClaim} and \texttt{AnalogyClaim} with
LLM-backed oracles, Pramana provides the audit-replayable documentary
artifact rather than deterministic re-verification (see §2.5).

Pramana provides a documentary artifact each framework requires but does
not by itself establish substantive compliance. NYDFS's ``rational,
statistically significant'' standard requires that the model's
underwriting logic in fact be rational and statistically significant;
§1557's ``reasonable efforts'' standard requires that the deployer's
efforts in fact be reasonable. Pramana makes these properties auditable
by recording the inputs, sources, and outputs of each decision; whether
the recorded properties hold under audit is a substantive evaluation
outside the protocol's scope and inside the regulator's. The protocol is
necessary documentary infrastructure; substantive compliance also
requires that the model and the deployer's process hold up under that
audit. Whether audit-replayability (as distinct from deterministic
re-verification) satisfies the substantive evidentiary standards of any
cited regulation is an open legal question without settled precedent;
this paper maps structural alignment, not substantive sufficiency, and
notes the open question rather than claiming to resolve it.

\subsubsection{4.1 Consumer lending (CFPB)}\label{consumer-lending-cfpb}

CFPB Circular 2023-03 {[}3{]} addresses creditors using complex
algorithms including AI for credit decisions. The Equal Credit
Opportunity Act and Regulation B require lenders to provide specific
reasons for adverse action, such that ``the model rejected you'' does
not satisfy the requirement and the CFPB sample-form checkboxes cannot
substitute for the actual reason. A Pramana-instrumented lender
produces, per credit decision, a \texttt{ClaimAttestation} chain in
which each adverse-action factor is grounded in a specific source claim
with a re-executable \texttt{verify()} against the underlying data. With
this artifact attached to the notice, the consumer and a CFPB examiner
can recover the basis for the decision by re-running the verification
chain.

\subsubsection{4.2 Banking model risk (SR 11-7, OCC
2011-12)}\label{banking-model-risk-sr-11-7-occ-2011-12}

Federal Reserve SR 11-7 and OCC Bulletin 2011-12 {[}4{]} is the
supervisory guidance on model risk management for banks, predating large
language models by a decade and applying to AI-driven models by default.
SR 11-7 requires documentation ``sufficiently detailed so that parties
unfamiliar with a model can understand how the model operates, its
limitations, and its key assumptions.'' The SR 11-7 examination surface
is what a \texttt{ClaimAttestation} chain provides, since the chain
records, per model output, what was claimed, against what source, by
which model, when, and how verification was performed. A bank using a
Pramana-instrumented agent for credit underwriting, fraud scoring, or
trading-strategy generation produces the SR 11-7 audit trail as a side
effect of normal operation.

\subsubsection{4.3 Insurance underwriting (NYDFS, Colorado
DOI)}\label{insurance-underwriting-nydfs-colorado-doi}

NYDFS Insurance Circular Letter No.~7 {[}5{]} requires insurers to
demonstrate that AI-driven underwriting variables have ``a clear,
empirical, statistically significant, rational, and not unfairly
discriminatory relationship'' with the insured risk. The ``demonstrate''
requirement maps directly to the \texttt{InferenceClaim} primitive,
whose \texttt{verify()} operation replays the inference against the
named premises and returns a deterministic outcome rather than a model
verdict. An NYDFS examiner reviewing an algorithmic underwriting
decision can re-execute the inference chain and verify each step.
Colorado Division of Insurance Regulation 10-1-1 {[}6{]} requires
governance and risk-management framework documentation for insurers
using external consumer data, algorithms, and predictive models. The
\texttt{ClaimAttestation} audit chain provides the documentary artifact
this framework requires as the artifact of normal operation.

\subsubsection{4.4 Healthcare (HIPAA, Section 1557, FDA
SaMD)}\label{healthcare-hipaa-section-1557-fda-samd}

HIPAA §164.312(b) audit controls require ``examination of activity'' in
ePHI systems, a rule general enough that it applies to algorithmic
decision aids by default. HHS OCR's Section 1557 Final Rule (45 C.F.R.
§92.210) {[}7{]} regulates patient care decision support tools and
requires covered entities to make ``reasonable efforts'' to identify and
mitigate discrimination from such tools. The FDA
Software-as-a-Medical-Device framework requires ``verifiable
performance'' of AI-driven clinical software. Per-claim
\texttt{VerificationOutcome} records provide the documentary artifact
required by the §164.312(b) examination requirement, the §92.210
reasonable-efforts standard, and the SaMD verifiability bar.

\subsubsection{4.5 Employment screening (NYC Local Law 144, EEOC,
ADA/ADEA/Title
VII)}\label{employment-screening-nyc-local-law-144-eeoc-adaadeatitle-vii}

AI-driven hiring tools are the most-deployed application of automated
decision-making in US labor markets. Fuller et al. {[}35{]} document
that applicant-tracking systems screen out roughly 27 million qualified
US workers -- veterans, people with disabilities, caregivers returning
after employment gaps, formerly incarcerated, and immigrants with
non-traditional credentials -- by treating status proxies as
disqualifiers. Per-candidate evidence of what the system claimed and
against what source is what an affected applicant or an auditor needs to
contest a decision.

NYC Local Law 144 {[}36{]} requires any employer using an Automated
Employment Decision Tool to commission an independent bias audit
annually, publish the audit summary, and notify candidates ten business
days in advance. The implementing rule defines the audit as
selection-rate disparity testing by protected category. Conducting that
audit on a black-box system requires reconstructing per-candidate
scoring from logs after the fact; conducting it on a
Pramana-instrumented system means the per-candidate
\texttt{ClaimAttestation} chain is the audit substrate by construction.

EEOC v. iTutorGroup {[}37{]} established that pre-existing federal
employment discrimination statutes (ADEA, Title VII) reach AI hiring
tools without new legislation. Mobley v. Workday {[}38{]} extended
liability to the AI vendor itself as an ``agent'' of its employer
customers, conditionally certifying a nationwide ADEA collective action.
In discovery, plaintiffs may demand the vendor's claim-level reasoning
trace; a vendor without one is in a weaker position than one holding the
documentary artifact this protocol standardizes. Whether that artifact
substantively prevents discrimination is for the regulator and the court
to evaluate, per the necessary-versus-sufficient distinction at the top
of this section.

\subsubsection{4.6 European Union and US state
laws}\label{european-union-and-us-state-laws}

EU AI Act Articles 14 (human oversight) and 50 (transparency) {[}8{]}
take effect August 2, 2026, for high-risk systems. As of writing, the EU
AI Office has not published technical guidance on consent or
verification mechanisms for autonomous agent transactions, and the
compliance deadline binds before guidance is final. Pramana's per-claim
\texttt{ClaimAttestation} is the artifact an oversight reviewer can
inspect (Article 14), and the discoverable extension URI plus per-claim
verification log provides the disclosure surface Article 50 requires.
GDPR Recital 71 {[}9{]} establishes the right to explanation; an
artifactless system cannot satisfy it. Three US states have enacted
broader automated-decision laws with audit-trail obligations: the
Colorado AI Act, California's ADMT (Automated Decision-Making
Technology) regulations, and Texas TRAIGA (Texas Responsible AI
Governance Act). Each statute's obligation maps directly to a
\texttt{ClaimAttestation} chain. NYC Local Law 144's employment-specific
mandate is covered in §4.5.

\subsubsection{4.7 Worked example: NYDFS-relevant underwriting
decision}\label{worked-example-nydfs-relevant-underwriting-decision}

Consider an autonomous underwriting agent processing an application
against NYDFS Insurance Circular Letter No.~7 {[}5{]} and its
``rational, statistically significant relationship'' requirement. The
agent emits three Pramana-attested claims that together constitute the
decision artifact.

\textbf{Source claim.} A \texttt{MeasurementClaim} records the
applicant's structured features (debt-to-income ratio, FICO score,
claims history over three years) with \texttt{measurement\_source}
pointing at the applicant's structured-record URI. \texttt{verify()}
fetches the source record and confirms field-match against the declared
\texttt{measured\_value}. The oracle is the structured-record store,
which is deterministic.

\textbf{Decision claim.} An \texttt{InferenceClaim} records the
underwriting verdict
(\texttt{claim\ =\ "applicant\ qualifies\ for\ tier\ 2\ pricing"}), the
premises that warrant it
(\texttt{premises\ =\ {[}"DTI\ \textless{}\ 0.40",\ "FICO\ \textgreater{}=\ 700",\ "no\ claims\ in\ 3\ years"{]}}),
and the named rule
(\texttt{inference\_method\ =\ "tier\_2\_eligibility\_rule\_v1.2"}).
\texttt{verify(oracle)} replays the rule against the premises. When the
oracle is a deterministic rule engine, the outcome is bit-identically
reproducible. When the oracle is an LLM, the verify() inputs and outputs
are audit-replayable per the scope clarification at the top of this
section.

\textbf{Policy claim.} A \texttt{CitationClaim} records the underwriting
policy reference
(\texttt{source\_uri\ =\ "policy://underwriting/v1.2\#section-3"},
\texttt{source\_excerpt\ =\ "Tier\ 2\ pricing\ applies\ when\ DTI\ \textless{}\ 0.40\ AND\ FICO\ \textgreater{}=\ 700..."}).
\texttt{verify()} re-fetches the policy and checks excerpt presence;
\texttt{source\_digest} carries the SHA-256 of the policy text at
retrieval time so post-decision tampering is detectable (Section 2.5).

The \texttt{claim-attestation/v1} wire envelope wraps each attestation
with \texttt{verify\_endpoint\_hint} (CA-1), \texttt{sla\_window\_ms}
(CA-2), and \texttt{source\_digest} plus \texttt{artifact\_signature}
(CA-3). An NYDFS examiner reviewing this decision queries the audit
endpoint, retrieves the three attestations and their
\texttt{VerificationOutcome} records, and re-executes \texttt{verify()}
against the recorded sources. The audit trail records each step in
append-only form.

This is the documentary record NYDFS Circular Letter 7 requires for any
algorithmic underwriting decision. Whether
\texttt{tier\_2\_eligibility\_rule\_v1.2} is in fact ``rational and
statistically significant'' is the substantive evaluation the regulator
performs, not what Pramana establishes (per the
necessary-versus-sufficient distinction at the top of this section).

The three attestations and their \texttt{VerificationOutcome} records,
emitted end-to-end by the reference implementation using the field names
and \texttt{verify()} semantics described above, are committed at
\texttt{samples/nydfs\_underwriting\_walkthrough.json} in the companion
repository; the script that produces them is at
\texttt{samples/build\_nydfs\_walkthrough.py}. All three claims verify
against in-memory mocks that stand in for the structured-record store,
the named rule engine, and the policy document store; in production
those slot in as the same dependency-injection points the unit tests
exercise.

Section 5 reports an exploratory empirical pilot that characterizes the
alternative pattern Pramana is designed to replace.

\begin{center}\rule{0.5\linewidth}{0.5pt}\end{center}

\subsection{5. Supplementary empirical
pilot}\label{supplementary-empirical-pilot}

The structural argument in Sections 1 through 4 motivates Pramana as the
wire format for re-verifiable claim artifacts. This section reports an
exploratory pilot characterizing how well the dominant alternative (an
LLM scoring another LLM's output) performs on code generation, the
most-studied LLM-as-judge surface. The pilot uses n=100 problems
totaling approximately 1.3M tokens across 2,275 reviewer calls (roughly
918K input plus 373K output tokens, computed from the JSONL logs in the
companion repository). It is not a controlled comparison: the three
ensemble configurations vary multiple factors at once (ensemble size,
capability tier, family composition, prompt-tuning), and the
prompt-selection set overlaps deterministically with the main-run
problem set. Findings are descriptive observations, not mechanism
claims. The pilot does not validate Pramana on its own; the structural
argument in Sections 1 through 4 and the formal verification in Section
6 do that.

\subsubsection{5.1 Design}\label{design}

MBPP (Mostly Basic Python Problems) {[}15{]} and HumanEval {[}16{]}
canonical solutions are mutated by AST-level operators (operator-swap,
off-by-one, edge-case-drop, method-confusion, type-confusion) to produce
quiet buggy code that is parseable, runnable, and fails the supplied
test suite. Three reviewer-ensemble configurations evaluate each (clean,
buggy) pair under three aggregation rules (unanimous, majority, any):
same-model (three independent Haiku 4.5 calls at temperature 0),
same-family (Haiku 4.5 + Sonnet 4.6 + Opus 4.7, all Anthropic), and
cross-family (Haiku 4.5 + GPT-4o-mini + Llama-3.3-70B). Prompt-variant
selection methodology is in the companion repository. The same-model
configuration doubles as an implicit single-model zero-shot baseline: at
temperature 0 the three Haiku calls produce identical verdicts, so its
detection and FPR are also the single-Haiku numbers. The ensemble
configurations are evaluated against this baseline.

\subsubsection{5.2 Strongest observation: the dataset-curation
effect}\label{strongest-observation-the-dataset-curation-effect}

Under the same ensemble configuration (same-family majority), the same
prompt, and the same model versions, raw FPR on HumanEval differs from
MBPP by roughly 40 percentage points (Table 1, Figure 1). The delta is
consistent with reference-solution quality contributing significantly:
MBPP canonical solutions include latent defects that reviewers correctly
flag but that the test suite does not exercise. Differential
prompt-overfit across the two corpora during the selection phase is an
alternative candidate driver that this design cannot rule out, and the
single-rater adjudication cannot establish that any one factor
dominates. Isolating drivers would require held-out replication on
disjoint problems and multi-rater adjudication. Reported
reviewer-ensemble FPR depends nontrivially on the reference-solution
distribution, so single-corpus FPR numbers are misleading without
disclosure of the test-suite quality. Per-case adjudication of
clean-flagged-buggy cases is in the companion repository; the
adjudication shows that what looks like a high false-positive rate on
MBPP is partly reviewers correctly catching latent defects the tests
miss.

\textbf{Table 1.} Per-corpus detection and FPR, majority rule, with 95\%
bootstrap CIs (10,000 resamples; n=40 per HumanEval cell, n=60 per MBPP
cell, per buggy/clean label). Numbers are subject to the optimistic-bias
caveat above.

{\def\LTcaptype{none} % do not increment counter
\begin{longtable}[]{@{}llll@{}}
\toprule\noalign{}
Ensemble & Corpus & Detection & FPR \\
\midrule\noalign{}
\endhead
\bottomrule\noalign{}
\endlastfoot
same-model & HumanEval & 97.5\% {[}92.5, 100.0{]} & 27.5\% {[}15.0,
42.5{]} \\
same-model & MBPP & 98.3\% {[}95.0, 100.0{]} & 66.7\% {[}55.0,
78.3{]} \\
same-family & HumanEval & 100.0\% {[}100.0, 100.0{]} & 5.0\% {[}0.0,
12.5{]} \\
same-family & MBPP & 96.7\% {[}91.7, 100.0{]} & 41.7\% {[}30.0,
53.3{]} \\
cross-family & HumanEval & 97.5\% {[}92.5, 100.0{]} & 27.5\% {[}15.0,
42.5{]} \\
cross-family & MBPP & 98.3\% {[}95.0, 100.0{]} & 60.0\% {[}48.3,
71.7{]} \\
\end{longtable}
}

\begin{figure}
\centering
\includegraphics[width=0.85\linewidth,height=\textheight,keepaspectratio,alt={Figure 1: Detection and FPR by ensemble structure, faceted by source corpus, majority rule. The same ensemble running the same prompt against the same model versions produces sharply different FPR on MBPP vs HumanEval; reference-solution quality and per-corpus prompt-overfit are both candidate drivers that this design cannot disentangle.}]{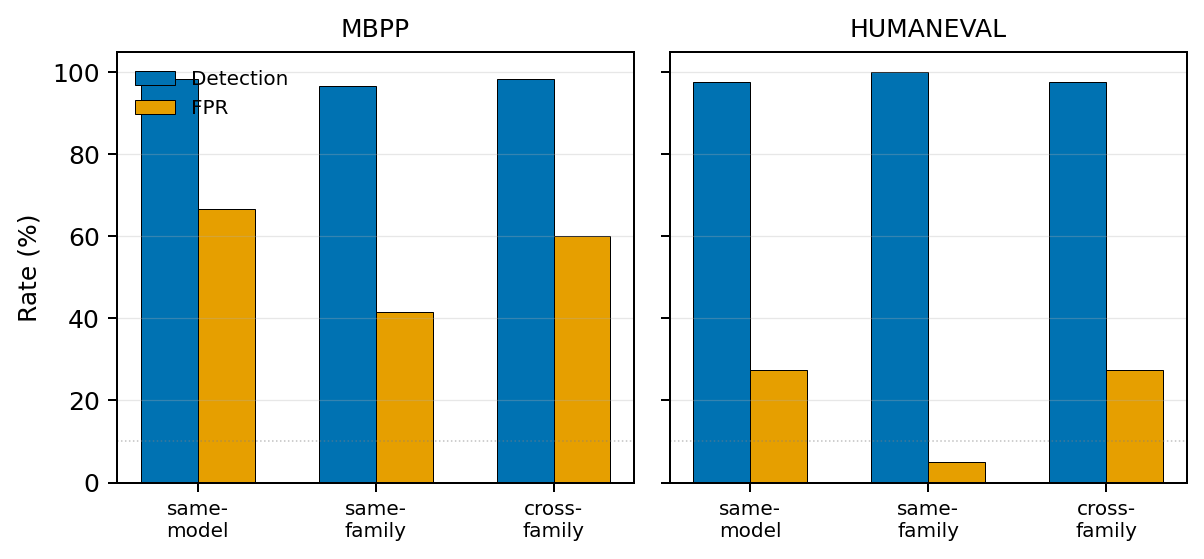}
\caption{Figure 1: Detection and FPR by ensemble structure, faceted by
source corpus, majority rule. The same ensemble running the same prompt
against the same model versions produces sharply different FPR on MBPP
vs HumanEval; reference-solution quality and per-corpus prompt-overfit
are both candidate drivers that this design cannot disentangle.}
\end{figure}

\subsubsection{5.3 Configuration-shift
observations}\label{configuration-shift-observations}

Detection and FPR per ensemble cell are in Table 2 (bootstrap 95\% CIs
over 10,000 resamples). Table 2's raw FPR estimates are subject to
optimistic bias from prompt-selection overlap with the evaluation set
(Section 8). The paired McNemar comparisons in Table 3 are reported as
descriptive observations of configuration differences. The prompt-pilot
optimized adjusted FPR for a single Haiku reviewer, so the leakage is
concentrated in Haiku-weighted configurations rather than symmetric
across them; the paired contrasts are therefore not strictly
bias-immune, and the p-values measure paired disagreement counts under
asymmetric leakage rather than treatment-effect significance.

\textbf{Table 2.} Detection and FPR per ensemble cell. n\_buggy = 100,
n\_clean = 100.

{\def\LTcaptype{none} % do not increment counter
\begin{longtable}[]{@{}llll@{}}
\toprule\noalign{}
Ensemble & Rule & Detection & FPR \\
\midrule\noalign{}
\endhead
\bottomrule\noalign{}
\endlastfoot
same-model & all rules & 98.0\% {[}95.0, 100.0{]} & 51.0\% {[}41.0,
61.0{]} \\
same-family & unanimous & 88.0\% {[}81.0, 94.0{]} & 10.0\% {[}5.0,
16.0{]} \\
same-family & majority & 98.0\% {[}95.0, 100.0{]} & 27.0\% {[}19.0,
36.0{]} \\
same-family & any & 99.0\% {[}97.0, 100.0{]} & 54.0\% {[}44.0,
64.0{]} \\
cross-family & unanimous & 83.0\% {[}75.0, 90.0{]} & 25.0\% {[}17.0,
34.0{]} \\
cross-family & majority & 98.0\% {[}95.0, 100.0{]} & 47.0\% {[}37.0,
57.0{]} \\
cross-family & any & 100.0\% {[}100.0, 100.0{]} & 79.0\% {[}71.0,
87.0{]} \\
\end{longtable}
}

\begin{figure}
\centering
\includegraphics[width=0.85\linewidth,height=\textheight,keepaspectratio,alt={Figure 2: Detection vs FPR by ensemble structure and aggregation rule, with 95\% bootstrap CIs.}]{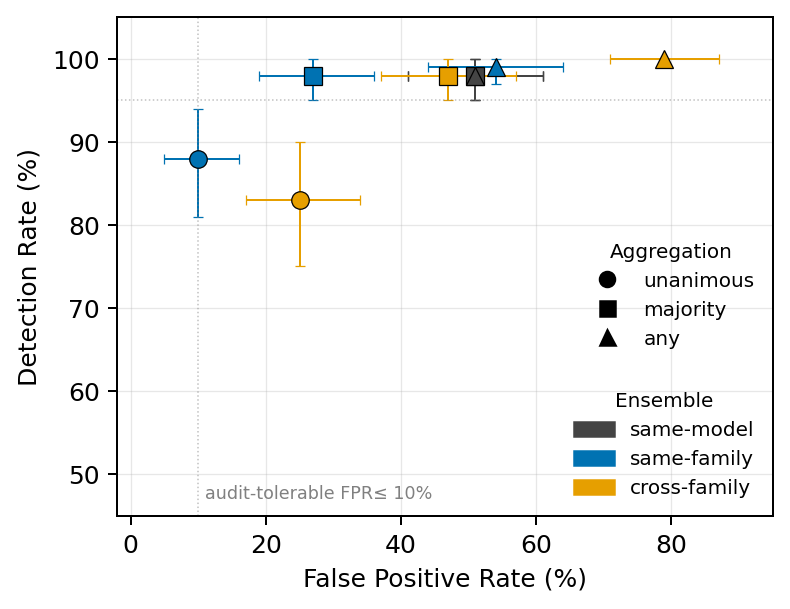}
\caption{Figure 2: Detection vs FPR by ensemble structure and
aggregation rule, with 95\% bootstrap CIs.}
\end{figure}

Under the majority rule, McNemar's exact paired comparison identifies
two configuration-shift effects at p\textless0.0001 (Table 3, Figure 3).
The 3-Haiku-at-temperature-0 configuration produces three identical
verdicts (collapsing to a single effective vote), and shifting to a
3-distinct-model same-family configuration flips 24 clean cases from
flagged-buggy to flagged-clean with zero shifts in the reverse
direction. Shifting from the same-family triple to the cross-provider
triple flips 22 cases in the opposite direction. The data does not
isolate the treatment effect: ensemble size, capability tier, family
composition, and prompt-tuning vary together. We report the McNemar
shifts as descriptive observations rather than diversity or scaling
claims. The p-values measure paired-disagreement counts on
configurations that vary multiple factors at once; readers should
interpret them as evidence that the configurations differ, not as
evidence for any specific causal mechanism. The same-model configuration
is documented for what it tests (production-reproducibility-mode T=0
ensembling) and does not speak to self-consistency, which requires
T\textgreater0 to sample diverse reasoning paths.

Two descriptive comparisons in the pilot hold most of their variables
constant. (a) The corpus split in Table 1 holds the ensemble constant
(same-family majority) and varies the evaluation set, isolating
reference-solution-quality and other inter-corpus differences as
candidate drivers. (b) The implicit single-Haiku baseline (the
same-model row) against the family-scaled ensemble \{Haiku, Sonnet,
Opus\} varies ensemble size and capability tier together but holds
family and prompt constant. Comparisons that vary family composition
independently of capability tier are not isolated by this design.

\textbf{Table 3.} McNemar's paired comparisons, majority rule.
Significant cells (p \textless{} 0.05) in bold. The p-values reflect the
direction of paired disagreement, not isolated treatment effects: each
comparison varies ensemble size, capability tier, family composition,
and prompt-tuning simultaneously. Readers should interpret the cells as
evidence that the configurations differ in their labeling, not as
evidence for any specific causal mechanism.

{\def\LTcaptype{none} % do not increment counter
\begin{longtable}[]{@{}
  >{\raggedright\arraybackslash}p{(\linewidth - 8\tabcolsep) * \real{0.2000}}
  >{\raggedright\arraybackslash}p{(\linewidth - 8\tabcolsep) * \real{0.2000}}
  >{\raggedright\arraybackslash}p{(\linewidth - 8\tabcolsep) * \real{0.2000}}
  >{\raggedright\arraybackslash}p{(\linewidth - 8\tabcolsep) * \real{0.2000}}
  >{\raggedright\arraybackslash}p{(\linewidth - 8\tabcolsep) * \real{0.2000}}@{}}
\toprule\noalign{}
\begin{minipage}[b]{\linewidth}\raggedright
Slice
\end{minipage} & \begin{minipage}[b]{\linewidth}\raggedright
Comparison
\end{minipage} & \begin{minipage}[b]{\linewidth}\raggedright
b
\end{minipage} & \begin{minipage}[b]{\linewidth}\raggedright
c
\end{minipage} & \begin{minipage}[b]{\linewidth}\raggedright
p (exact)
\end{minipage} \\
\midrule\noalign{}
\endhead
\bottomrule\noalign{}
\endlastfoot
buggy (detection) & all three pairwise & \textless=1 & \textless=1 &
1.000 \\
clean (FPR) & \textbf{same-model vs same-family} & \textbf{24} &
\textbf{0} & \textbf{\textless0.0001} \\
clean (FPR) & \textbf{same-family vs cross-family} & \textbf{2} &
\textbf{22} & \textbf{\textless0.0001} \\
clean (FPR) & same-model vs cross-family & 5 & 1 & 0.219 \\
\end{longtable}
}

\begin{figure}
\centering
\includegraphics[width=0.85\linewidth,height=\textheight,keepaspectratio,alt={Figure 3: McNemar discordant-cell counts on the clean slice for the two significant pairwise comparisons. Each comparison varies multiple factors at once (ensemble size, capability tier, family composition); the shifts are reported as configuration differences rather than attributable to a single treatment effect.}]{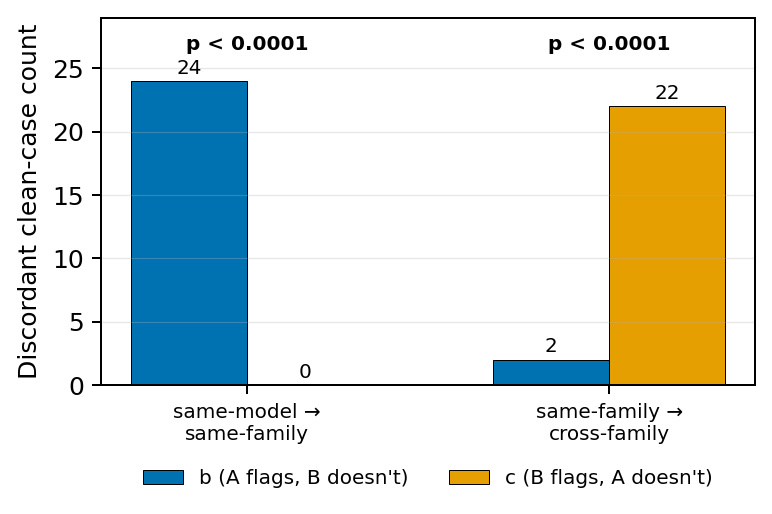}
\caption{Figure 3: McNemar discordant-cell counts on the clean slice for
the two significant pairwise comparisons. Each comparison varies
multiple factors at once (ensemble size, capability tier, family
composition); the shifts are reported as configuration differences
rather than attributable to a single treatment effect.}
\end{figure}

The raw program output from \texttt{metrics.py} against the run's JSONL
log reproduces the cells above:

\begin{verbatim}
=== Detection & FPR (point [95% bootstrap CI]) ===
  same-family    unanimous      88.0% [81.0, 94.0]     10.0% [ 5.0, 16.0]
  same-family    majority       98.0% [95.0, 100.0]    27.0% [19.0, 36.0]
  cross-family   majority       98.0% [95.0, 100.0]    47.0% [37.0, 57.0]

=== McNemar's paired comparisons (rule=majority, clean slice) ===
  same-model     vs same-family       24    0  22.04   0.0000
  same-family    vs cross-family       2   22  15.04   0.0000
\end{verbatim}

\subsubsection{5.4 Why the alternative fails
structurally}\label{why-the-alternative-fails-structurally}

The regress that §7.1 names structurally is observable in the pilot.
External ground truth from test suites is partial since unit tests miss
bugs. Improving the ground truth either reduces to writing better tests
(the original problem the reviewer was deployed to solve) or requires a
stronger judge with the same problem one level up. Three classes of
partial fix exist (importing external ground truth, applying calibration
regularization, removing the LLM from the judgment step entirely); the
first two push the problem one level up while still requiring the
labelled data they were meant to substitute for. Pramana takes the third
route for \texttt{MeasurementClaim} and \texttt{CitationClaim} in full,
and partially for \texttt{InferenceClaim} and \texttt{AnalogyClaim} when
the oracle is itself deterministic (see Section 2.5). The structural
escape is the contribution; the empirical pilot documents that the
alternative (probabilistic reviewer ensembles) faces the loop.

\subsubsection{5.5 Adversarial robustness
check}\label{adversarial-robustness-check}

Five programmatic adversarial transformations were applied to 30 sampled
buggy mutations; the same-family majority detection per transformation
is in Table 4 and Figure 4.

\textbf{Table 4.} Adversarial detection per transformation. Mean benign
baseline is 98.0\%.

{\def\LTcaptype{none} % do not increment counter
\begin{longtable}[]{@{}lll@{}}
\toprule\noalign{}
Transformation & Detection & \(\Delta\) vs benign \\
\midrule\noalign{}
\endhead
\bottomrule\noalign{}
\endlastfoot
misleading\_docstring & 96.7\% & −1.3 pp \\
defensive\_assertion & 93.3\% & −4.7 pp \\
rationalizing\_comment & 96.7\% & −1.3 pp \\
noisy\_helpers & 96.7\% & −1.3 pp \\
plausible\_rename & 96.7\% & −1.3 pp \\
\end{longtable}
}

\begin{figure}
\centering
\includegraphics[width=0.85\linewidth,height=\textheight,keepaspectratio,alt={Figure 4: Detection rate per adversarial transformation, same-family majority. The benign baseline (98.0\%) is shown as the dashed reference line. The largest drop is 4.7 pp (defensive\_assertion); the mean drop is 2.0 pp.}]{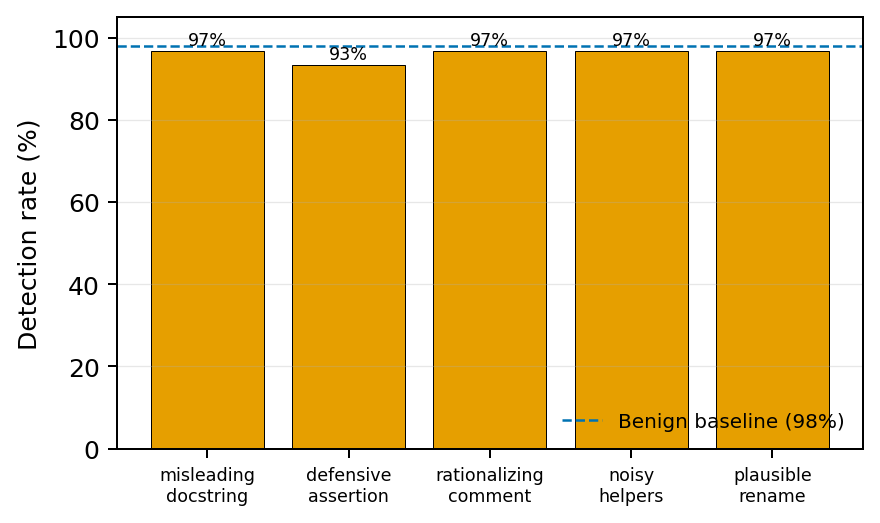}
\caption{Figure 4: Detection rate per adversarial transformation,
same-family majority. The benign baseline (98.0\%) is shown as the
dashed reference line. The largest drop is 4.7 pp
(defensive\_assertion); the mean drop is 2.0 pp.}
\end{figure}

\subsubsection{5.6 Scope and methodological
constraints}\label{scope-and-methodological-constraints}

The pilot uses code generation, which has objective binary ground truth
via unit tests; the regulated domains the protocol targets have ground
truth that is contested, probabilistic, or unavailable at decision time.
The pilot cannot directly simulate those conditions. The adjudication of
clean-flagged-buggy cases is single-rater, a methodological ceiling on
the adjusted-FPR numbers; the mathematical analyses of specific failure
modes (Vieta's formulas for reciprocal roots, triangle-in-semicircle
area, range arithmetic, cyclic-encoding identity) are externally
verifiable and invariant to rater bias. The adversarial check at n=30
per transformation surfaces detection drops of 1.3 to 4.7 percentage
points; this is exploratory, with effect sizes within sampling noise at
this sample size. A pilot in a regulated domain is the natural next
study. Such a pilot would implement a Pramana-instrumented agent against
a documented decision-rule set (mock underwriting against NYDFS-relevant
variables, mock diagnostic against §92.210- relevant EHR (electronic
health record) fields), generate \texttt{ClaimAttestation} chains for
representative decisions, and submit those artifacts to compliance
review against the specific regulatory framework's bar. The current
paper establishes the protocol design (Section 2), formal lifecycle
correctness (Section 6), the reference implementation, and the
structural argument that the artifact-shape regulators require is what
Pramana produces. Empirical regulatory-compliance validation is the next
paper in this programme.

The lifecycle properties that make this primitive correct under
concurrency, suppression, and disclosure are formalized next.

\begin{center}\rule{0.5\linewidth}{0.5pt}\end{center}

\subsection{6. Formal properties}\label{formal-properties}

\subsubsection{6.1 Specification
structure}\label{specification-structure}

The TLA+ specification at \texttt{specification/Pramana.tla} models four
lifecycle actions (\texttt{EmitClaim}, \texttt{VerifyClaim},
\texttt{SuppressClaim}, \texttt{ShowToPrincipal}) and tracks emission,
verification state, suppression, principal view, and an append-only
audit trail. State variables capture per-claim type, source, attester,
and verification outcome. The audit trail is a sequence of records, each
tagged with the action that produced it. The five named safety
invariants from Section 2.3 (P-1 through P-4, plus
SuppressionDisclosureDisjoint) are TLC \texttt{INVARIANTS}.

\subsubsection{6.2 Models and results}\label{models-and-results}

Three model configurations exhaustively check the invariants.

{\def\LTcaptype{none} % do not increment counter
\begin{longtable}[]{@{}
  >{\raggedright\arraybackslash}p{(\linewidth - 4\tabcolsep) * \real{0.3333}}
  >{\raggedright\arraybackslash}p{(\linewidth - 4\tabcolsep) * \real{0.3333}}
  >{\raggedright\arraybackslash}p{(\linewidth - 4\tabcolsep) * \real{0.3333}}@{}}
\toprule\noalign{}
\begin{minipage}[b]{\linewidth}\raggedright
Model
\end{minipage} & \begin{minipage}[b]{\linewidth}\raggedright
Distinct states
\end{minipage} & \begin{minipage}[b]{\linewidth}\raggedright
Result
\end{minipage} \\
\midrule\noalign{}
\endhead
\bottomrule\noalign{}
\endlastfoot
Lifecycle (P-1 through P-3) & 2,345 & 0 violations \\
Disclosure (P-1 through P-4, plus disjointness) & 4,601 & 0
violations \\
Concurrency (multi-actor) & 31,617 & 0 violations \\
\end{longtable}
}

The total across the three models is \textbf{38,563 distinct reachable
states with zero invariant violations}. The three-model split is the
standard TLA+ small-model pattern, in which focused minimal models
converge in seconds rather than one large model that does not. All three
use \texttt{SYMMETRY\ ModelSymmetry} over the agent, claim, source, and
verifier sets, with a \texttt{BoundedTrail} constraint capping
\texttt{Len(auditTrail)} at 6. The verification is reproducible via
\texttt{specification/run\_tlc.sh}. The raw TLC output captures the
per-model state counts:

\begin{verbatim}
[Lifecycle]   2349 states generated, 2345 distinct states found, 0 states left on queue.
[Disclosure]  7409 states generated, 4601 distinct states found, 0 states left on queue.
[Concurrency] 47689 states generated, 31617 distinct states found, 0 states left on queue.
\end{verbatim}

A supplementary partial-exhaustive run (3 claims, 2 agents, 2 verifiers,
1 principal) was explored to 166M+ distinct states with zero violations
before BFS (breadth-first search) exhausted compute.

\subsubsection{6.3 Wire-extension
invariants}\label{wire-extension-invariants}

The \texttt{claim-attestation/v1} wire extension adds three
deployment-grade invariants beyond the core lifecycle invariants: CA-1
(reachability), CA-2 (SLA-bound), and CA-3 (offline re-verifiability).
CA-1 and CA-3 are separately TLC-verified at 21 distinct states with
zero violations. CA-2 is a runtime invariant enforced by validators in
the reference implementation.

\subsubsection{6.4 Reference
implementation}\label{reference-implementation}

The Python reference implementation passes 84 tests (59 core and 25
extension): 14 primitive-type tests, 34 across the four
\texttt{verify()} implementations covering every outcome branch, 6
TLC-config tripwires preventing accidental drift between the spec and
the runnable model, and 5 Hypothesis property-based tests fuzzing
wire-format invariants (JSON round-trip, discriminator dispatch,
claim-type invariance, never-raises, populated \texttt{outcome.method}),
plus 25 claim-attestation extension tests. The suite runs in under one
second. Combined with TLC verification, this gives end-to-end
correctness coverage from the specification to the implementation.

Related work spans the empirical critique of LLM-as-judge,
provenance-grounded alternatives (including W3C Verifiable Credentials
and PROV), agent-loop systems, and regulatory scholarship.

\begin{center}\rule{0.5\linewidth}{0.5pt}\end{center}

\subsection{7. Related work}\label{related-work}

\subsubsection{7.1 Empirical critique of LLM-as-judge and
reviewer-ensemble
patterns}\label{empirical-critique-of-llm-as-judge-and-reviewer-ensemble-patterns}

Reviewer-ensemble evaluation inherits a measurement regress. Measuring
the accuracy of an LLM judge requires ground truth; the ground truth
from external test suites is partial because unit tests miss bugs;
improving the ground truth either reduces to the original problem the
reviewer was deployed to solve or requires a stronger judge that faces
the same problem one level up. The regress is structural, not a tuning
issue: any pipeline that uses an LLM to judge another LLM's output
inherits it. The empirical work surveyed below documents specific
failure modes that are downstream of this structural problem.

A growing body of empirical work shows that LLM judges and reviewer
ensembles do not perform the verification function they are deployed
for. Kim et al.~{[}17{]} observe that conditional error correlation
across 350+ models on two leaderboards is approximately 60\% (agreement
given both wrong), with correlation higher within a family and
persisting across providers when training corpora overlap, and
explicitly flag implications for the LLM-as-judge paradigm and
multi-agent systems. Li et al.~{[}18{]} document \emph{preference
leakage}: when a generator and a judge share lineage, the judge favors
the generator's outputs by margins larger than the well-known egocentric
bias. Zheng et al. {[}19{]}, Panickssery et al.~{[}20{]}, and Wang et
al.~{[}21{]} critique LLM-as-judge for offline benchmark scoring.
Stureborg et al. {[}22{]} show that LLM evaluators are systematically
inconsistent across runs. Maloyan et al.~{[}23{]} investigate
prompt-injection vulnerabilities in judge architectures. Nasr et
al.~{[}24{]} bypass twelve recent defense systems at greater than 90\%
success rate. Cemri et al.~{[}25{]} (MAST) taxonomize 14 multi-agent
failure modes and identify a verification-deficit category without
proposing a wire-format remedy. None of these works names the production
pattern of generator-plus-reviewer-ensemble-as-safety-gate as an
anti-pattern, characterizes it at the protocol layer, and proposes a
constructive wire-format alternative. The empirical work demonstrates
the problem; Pramana proposes the primitive.

\subsubsection{7.2 Provenance-grounded
alternatives}\label{provenance-grounded-alternatives}

Citation-Grounded Code Comprehension {[}26{]} introduced mechanical
citation verification and argued for architectural prevention superior
to post-hoc detection. Cited but Not Verified {[}27{]} documents the gap
between citations produced and citations that resolve to claimed
content. SelfCheckGPT {[}28{]} approaches the same problem through
within-model sampling, which is closer in spirit to reviewer-ensembles
than to Pramana and is complementary rather than competing. The survey
{[}29{]} confirms that no protocol in active use defines a typed
attestation surface for agent claims.

\textbf{W3C Verifiable Credentials.} The Verifiable Credentials Data
Model {[}39{]} standardizes typed identity and attribute attestations
with cryptographic verifiability. VC's design centre is the
identity-attestation use case (an issuer asserts a property of a
subject); its data model does not type agent claims by epistemic ground,
does not specify a \texttt{verify()} operation against a recorded
source, and does not address the LLM-oracle determinism partition.
Pramana adopts VC's architectural pattern of typed wire attestations and
extends it with the four-way epistemic taxonomy, the \texttt{verify()}
contract, and the wire-extension invariants specific to autonomous-agent
claim verification.

\textbf{General-purpose provenance models.} The W3C PROV family,
comprising the PROV Data Model PROV-DM {[}33{]} and the OWL (Web
Ontology Language) ontology PROV-O {[}34{]}, is the established standard
for domain-agnostic provenance, modelling entities, activities, and
agents along with their causal relations. PROV is deliberately a
meta-model: it describes how to record that an artifact was derived from
another but leaves the substrate-specific verification semantics to the
consuming domain. Pramana composes with PROV at a different layer rather
than competing with it. A Pramana \texttt{ClaimAttestation}, viewed
through a PROV lens, is a typed \texttt{prov:Entity} derived from a
\texttt{prov:Activity} (the agent's reasoning step) acting on a source
\texttt{prov:Entity}, and a successful \texttt{verify()} outcome is a
\texttt{prov:wasDerivedFrom} edge that a third party can independently
reconstruct. What Pramana adds on top of PROV is (i) a typed taxonomy of
four claim types with type-specific verification semantics rather than a
single generic derivation edge, (ii) wire-format invariants (CA-1, CA-2,
CA-3) that make those derivations checkable end-to-end across agents,
and (iii) a \texttt{verify()} contract whose output is deterministic
when paired with a deterministic oracle and offline re-runnable rather
than a curator-authored provenance record. A Pramana-conformant agent
could emit PROV-O serialisations of its audit trail; that mapping is
straightforward and is left as future work.

\subsubsection{7.3 Agent-loop and self-improvement
systems}\label{agent-loop-and-self-improvement-systems}

A separate line of recent work pairs LLMs with deterministic verifiers
in iterative loops. FunSearch {[}30{]} and AlphaEvolve {[}31{]} both use
a candidate-generator plus deterministic-evaluator architecture to make
new mathematical and algorithmic discoveries. The generator proposes
candidates and the evaluator verifies each candidate against a fitness
function or test suite that does not invoke another LLM, such that the
loop's correctness rests on the evaluator and not on the generator.
Google's AI co-scientist {[}42{]} extends the pattern to multi-agent
biomedical research, pairing specialized generation and evaluation
agents in tournament-style refinement loops that produce structured
hypothesis artifacts at every step. None of these systems define a
shared wire format across vendors or across agents, which is the gap
Pramana fills. Pramana is the wire-format analogue of this architecture
for agent outputs: every claim is paired with a \texttt{verify()}
against the recorded source, and no probabilistic judge participates in
the verification step.

\subsubsection{7.4 Regulatory and policy
work}\label{regulatory-and-policy-work}

Regulatory and policy work on algorithmic auditing has matured
separately from the wire-format literature. The frameworks cited in
Section 4 reflect a regulator-side convergence on per-decision
documentary records as the audit substrate. Dell'Acqua et al.~{[}32{]}
characterize the ``jagged frontier'' of AI capability that motivates
per-claim-type verification rather than uniform model-level treatment.
Mökander et al.~{[}40{]} survey ethics-based auditing methodology and
distinguish process-based from artifact-based audit substrates;
Pramana's per-decision record sits in the latter category. Kroll et
al.~{[}41{]} develop the ``accountable algorithms'' framing from the
legal-scholarship side, arguing that verifiable algorithmic records are
a necessary input to substantive due-process review while emphasizing
that structural documentation alone does not adjudicate substantive
sufficiency. Whether structural documentation in the Pramana sense
suffices for substantive evidentiary standards (and the related question
of vendor versus deployer liability raised in Mobley v. Workday
{[}38{]}) remains an open research area outside this paper's scope.

\begin{center}\rule{0.5\linewidth}{0.5pt}\end{center}

\subsection{8. Limitations and future
work}\label{limitations-and-future-work}

\begin{itemize}
\tightlist
\item
  \textbf{Pilot scope and domain mismatch.} Section 5 uses code
  generation, which has objective binary ground truth via unit tests;
  the regulated domains the protocol targets have ground truth that is
  contested, probabilistic, or unavailable at decision time. The pilot
  cannot directly simulate those conditions. Per-domain validation in
  regulated domains is open work and the natural next study.
\item
  \textbf{Sample size and confounding.} The pilot uses n=100 problems
  and configuration-shift comparisons confound multiple variables at
  once (ensemble size, capability tier, family composition,
  prompt-tuning). We report McNemar shifts as descriptive observations
  rather than mechanism claims. A larger study with held-out problem
  sets, a T\textgreater0 self-consistency control, and factorial
  separation would substantiate quantitative claims.
\item
  \textbf{Prompt-pilot and main-run overlap.} Prompt-variant selection
  drew from the same first-N positions of each dataset that the main run
  later evaluates. The selection optimized adjusted FPR on a single
  Haiku reviewer while the main run measures 7-slot ensemble FPR, so the
  leakage's effect on ensemble-level estimates is bounded by the
  imperfect coupling between single-reviewer and ensemble-reviewer FPR.
  The leakage is concentrated in Haiku-weighted ensemble configurations
  (same-model uses three Haiku calls; same-family and cross-family
  include one Haiku call each), so the paired McNemar comparisons across
  configurations operate under asymmetric leakage rather than symmetric
  bias; the §5.3 reporting reflects this. Held-out replication is open
  work.
\item
  \textbf{Single-rater adjudication.} Adjudication of
  clean-flagged-buggy cases is single-rater, a methodological ceiling on
  the adjusted-FPR numbers. The mathematical analyses of specific
  failure modes are externally verifiable and invariant to rater bias;
  the percentage splits are first-rater estimates. Inter-rater
  reliability measures such as Cohen's kappa are not reported, as
  adjudication used a single rater; a multi-rater replication with
  reported kappa is the natural next-study extension.
\item
  \textbf{Adversarial source fabrication.} A \texttt{CitationClaim}
  whose source resolves to a fake source the adversary controls is
  observable under Pramana but not prevented. Source-attestation
  primitives compose with Pramana via the \texttt{source\_uri}
  integration point and are out of scope for v1.
\item
  \textbf{AnalogyClaim verification semantics.} The v1 wire format
  treats \texttt{AnalogyClaim} VERIFIED outcomes uniformly whether or
  not a \texttt{similarity\_score} threshold was applied (§2.2).
  Regulated deployments should configure their emitters to require
  \texttt{similarity\_score}; a future revision could introduce a
  distinct \texttt{VERIFIED\_UNTHRESHOLDED} status so the distinction is
  visible at the wire-format inspection layer rather than only via
  deployer configuration.
\item
  \textbf{TLA+ model size.} Exhaustive verification uses minimal
  symmetry-reduced models (2 claims, BoundedTrail = 6). A supplementary
  3-claim, 2-agent, 2-verifier run reached 166M+ states with zero
  violations before BFS exhausted compute. Full inductive proof for
  arbitrary deployment scales is deferred to a TLAPS (TLA+ Proof System)
  treatment.
\item
  \textbf{Capability attestation and multi-agent composition.}
  Third-party capability attestation is absorbed by Yathartha {[}11{]}
  as a sibling primitive. Multi-agent claim composition (an
  \texttt{InferenceClaim} depending on \texttt{MeasurementClaim}s from
  two other agents) is a v2 extension. The TLA+ spec models single-agent
  lifecycles.
\end{itemize}

\begin{center}\rule{0.5\linewidth}{0.5pt}\end{center}

\subsection{References}\label{references}

\begin{enumerate}
\def\labelenumi{\arabic{enumi}.}
\tightlist
\item
  \emph{Agent2Agent (A2A) Protocol Specification.} a2a-protocol.org.
\item
  \emph{Model Context Protocol (MCP) Specification.}
  modelcontextprotocol.io.
\item
  Consumer Financial Protection Bureau. (2023, September 19).
  \emph{Adverse Action Notification Requirements and the Proper Use of
  the CFPB's Sample Forms Provided in Regulation B.} Circular 2023-03.
\item
  Board of Governors of the Federal Reserve System \& Office of the
  Comptroller of the Currency. (2011, April 4). \emph{Supervisory
  Guidance on Model Risk Management.} SR 11-7 / OCC Bulletin 2011-12.
\item
  New York State Department of Financial Services. (2024, July 11).
  \emph{Use of Artificial Intelligence Systems and External Consumer
  Data and Information Sources in Insurance Underwriting and Pricing.}
  Insurance Circular Letter No.~7 (2024).
\item
  Colorado Division of Insurance. (2023, September 21; effective
  November 14, 2023; expanded October 15, 2025). \emph{Governance and
  Risk Management Framework Requirements for Insurers' Use of External
  Consumer Data and Information Sources, Algorithms, and Predictive
  Models.} Regulation 10-1-1.
\item
  U.S. Department of Health and Human Services, Office for Civil Rights,
  \& Centers for Medicare \& Medicaid Services. (2024, May 6).
  \emph{Nondiscrimination in Health Programs and Activities}, Final
  Rule. 89 Fed. Reg. 28822 (codified at 45 C.F.R. pt.~92; §92.210
  governs patient care decision support tools).
\item
  \emph{EU Artificial Intelligence Act (Regulation (EU) 2024/1689),
  Articles 14, 50.} Effective August 2, 2026.
\item
  \emph{General Data Protection Regulation (GDPR), Recital 71.}
\item
  Kadaboina, R. K. (2026). \emph{Anumati: Proof of Adherence as a Formal
  Consent Model for Autonomous Agent Protocols.} arXiv:2604.16524.
\item
  Kadaboina, R. K. (2026). \emph{Yathartha: A Protocol-Layer Treatment
  of Jagged Intelligence in Autonomous Agent Networks.} Zenodo DOI
  10.5281/zenodo.19659633.
\item
  Kadaboina, R. K. (2026). \emph{Phala: Principal-Declared Welfare
  Feedback for Autonomous Agent Networks.} Zenodo DOI
  10.5281/zenodo.19625612.
\item
  Kadaboina, R. K. (2026). \emph{Pratyahara: A Neural Tissue Defense
  Model for Detecting Compromised Agents in Multi-Agent Networks.}
  Specification name: NERVE. Zenodo DOI 10.5281/zenodo.19628589.
\item
  Kadaboina, R. K. (2026). \emph{Sauvidya: An Accessibility Protocol for
  Agent-to-Principal Interaction in Autonomous Agent Networks.}
  Specification name: PACE. Zenodo DOI 10.5281/zenodo.19633139.
\item
  Austin, J., et al.~(2021). \emph{Program Synthesis with Large Language
  Models.} arXiv:2108.07732. (MBPP dataset.)
\item
  Chen, M., et al.~(2021). \emph{Evaluating Large Language Models
  Trained on Code.} arXiv:2107.03374. (HumanEval dataset.)
\item
  Kim, E., Garg, A., Peng, K., \& Garg, N. (2025). \emph{Correlated
  Errors in Large Language Models.} arXiv:2506.07962.
\item
  Li, D., et al.~(2025). \emph{Preference Leakage: A Contamination
  Problem in LLM-as-a-Judge.} arXiv:2502.01534.
\item
  Zheng, L., et al.~(2023). \emph{Judging LLM-as-a-Judge with MT-Bench
  and Chatbot Arena.} arXiv:2306.05685.
\item
  Panickssery, A., Bowman, S. R., \& Feng, S. (2024). \emph{LLM
  Evaluators Recognize and Favor Their Own Generations.}
  arXiv:2404.13076.
\item
  Wang, P., et al.~(2023). \emph{Large Language Models Are Not Fair
  Evaluators.} arXiv:2305.17926.
\item
  Stureborg, R., Alikaniotis, D., \& Suhara, Y. (2024). \emph{Large
  Language Models Are Inconsistent and Biased Evaluators.}
  arXiv:2405.01724.
\item
  Maloyan, N., Ashinov, B., \& Namiot, D. (2025). \emph{Investigating
  the Vulnerability of LLM-as-a-Judge Architectures to Prompt-Injection
  Attacks.} arXiv:2505.13348.
\item
  Nasr, M., et al.~(2025). \emph{The Attacker Moves Second: Stronger
  Adaptive Attacks Bypass Defenses Against LLM Jailbreaks and Prompt
  Injections.} arXiv:2510.09023.
\item
  Cemri, M., et al.~(2025). \emph{Why Do Multi-Agent LLM Systems Fail?}
  (MAST). arXiv:2503.13657.
\item
  Arafat, J. (2025). \emph{Citation-Grounded Code Comprehension.}
  arXiv:2512.12117.
\item
  Onweller, H., et al.~(2026). \emph{Cited but Not Verified: Parsing and
  Evaluating Source Attribution in LLM Deep Research Agents.}
  arXiv:2605.06635.
\item
  Manakul, P., Liusie, A., \& Gales, M. J. F. (2023).
  \emph{SelfCheckGPT: Zero-Resource Black-Box Hallucination Detection
  for Generative Large Language Models.} arXiv:2303.08896.
\item
  Yang, Y., et al.~(2025). \emph{A Survey of AI Agent Protocols.}
  arXiv:2504.16736.
\item
  Romera-Paredes, B., et al.~(2024). \emph{FunSearch: Mathematical
  Discoveries from Program Search with Large Language Models.} Nature
  625, 468-475.
\item
  Novikov, A., et al.~(2025). \emph{AlphaEvolve: A Coding Agent for
  Scientific and Algorithmic Discovery.} arXiv:2506.13131.
\item
  Dell'Acqua, F., McFowland III, E., Mollick, E., Lifshitz-Assaf, H.,
  Kellogg, K. C., Rajendran, S., Krayer, L., Candelon, F., \& Lakhani,
  K. R. (2023). \emph{Navigating the Jagged Technological Frontier:
  Field Experimental Evidence of the Effects of Artificial Intelligence
  on Knowledge Worker Productivity and Quality.} Harvard Business School
  Working Paper No.~24-013. Forthcoming in Organization Science.
\item
  Moreau, L., \& Missier, P. (eds.). (2013, April 30). \emph{PROV-DM:
  The PROV Data Model.} W3C Recommendation.
  \url{https://www.w3.org/TR/prov-dm/}
\item
  Lebo, T., Sahoo, S., \& McGuinness, D. (eds.). (2013, April 30).
  \emph{PROV-O: The PROV Ontology.} W3C Recommendation.
  \url{https://www.w3.org/TR/prov-o/}
\item
  Fuller, J., Raman, M., Sage-Gavin, E., \& Hines, K. (2021, September).
  \emph{Hidden Workers: Untapped Talent.} Harvard Business School
  Project on Managing the Future of Work, in collaboration with
  Accenture.
\item
  New York City Department of Consumer and Worker Protection. (2023).
  \emph{Automated Employment Decision Tools: Final Rule.} 6 RCNY § 5-300
  et seq. Implementing Local Law 144 of 2021; enforcement effective July
  5, 2023.
\item
  EEOC v. iTutorGroup, Inc., No.~1:22-cv-02565 (E.D.N.Y.). Consent
  decree filed August 9, 2023; settlement \$365,000.
\item
  Mobley v. Workday, Inc., No.~3:23-cv-00770 (N.D. Cal.). Order on
  motion to dismiss, July 12, 2024 (vendor liability as ``agent'' under
  Title VII, ADEA, ADA); collective action conditionally certified May
  16, 2025 (ADEA claims).
\item
  W3C. (2025). \emph{Verifiable Credentials Data Model v2.0.} W3C
  Recommendation. \url{https://www.w3.org/TR/vc-data-model-2.0/}
\item
  Mökander, J., Morley, J., Taddeo, M., \& Floridi, L. (2021, July).
  \emph{Ethics-Based Auditing of Automated Decision-Making Systems:
  Nature, Scope, and Limitations.} Science and Engineering Ethics,
  27(4). DOI 10.1007/s11948-021-00319-4.
\item
  Kroll, J. A., Huey, J., Barocas, S., Felten, E. W., Reidenberg, J. R.,
  Robinson, D. G., \& Yu, H. (2017). Accountable Algorithms.
  \emph{University of Pennsylvania Law Review}, 165(3), 633-705.
\item
  Gottweis, J., et al.~(2025). \emph{Towards an AI co-scientist.}
  arXiv:2502.18864.
\end{enumerate}

\begin{center}\rule{0.5\linewidth}{0.5pt}\end{center}

\subsection{Appendix: artifacts and
replication}\label{appendix-artifacts-and-replication}

The Pramana reference implementation, TLA+ specifications, A2A and MCP
discovery manifests, claim-attestation wire extension, empirical pilot
scripts and raw API responses, the PREREGISTRATION.md artifact, and
per-case adjudication files are at
\url{https://github.com/ravikiran438/pramana-attestation} under the
Apache 2.0 license. The TLC reproducer script
(\texttt{specification/run\_tlc.sh}) regenerates the state counts in
Section 6.2; full per-type schemas, the wire envelope specification, and
the empirical pilot reproducer scripts are in the same repository.

\end{document}